\documentclass[review]{elsarticle}

\usepackage{enumitem}
\usepackage{amsmath}
\usepackage{graphicx}
\usepackage{caption}
\usepackage{subcaption}
\usepackage{amsmath}
\usepackage{enumitem}
\usepackage{color}
\usepackage{graphicx}
\journal{Integration, the VLSI Journal}

\begin{document}

\begin{frontmatter}

\title{Hardware design of \emph{LIF with Latency} neuron model with memristive STDP synapses}

\author[label1]{Simone Acciarito}
\author[label1]{Gian Carlo Cardarilli\corref{mycorrespondingauthor}} 
\author[label1]{Alessandro Cristini}
\author[label1]{Luca Di Nunzio}
\author[label1]{Rocco Fazzolari} 
\author[label1]{Gaurav Mani Khanal}
\author[label1]{Marco Re}
\author[label1,label3]{Gianluca Susi}

\address[label1]{Department of Electronics Engineering. University of Rome ``Tor Vergata'', Rome, Italy}
\address[label3]{Laboratory of Cognitive and Computational Neuroscience (UCM-UPM), Center for Biomedical Technology. Technical University of Madrid, Madrid, Spain}

\begin{abstract}

In this paper, the hardware implementation of a neuromorphic system is presented. This system is composed of a Leaky Integrate-and-Fire with Latency (LIFL) neuron and a Spike-Timing Dependent Plasticity (STDP) synapse. LIFL neuron model allows to encode more information than the common Integrate-and-Fire models, typically considered for neuromorphic implementations. In our system LIFL neuron is implemented using CMOS circuits while memristor is used for the implementation of the STDP synapse.
A description of the entire circuit is provided.
Finally, the capabilities of the proposed architecture have been evaluated by simulating a motif composed of three neurons and two synapses.
The simulation results confirm the validity of the proposed system and its suitability for the design of  more complex spiking neural networks.

% The authors have been presented new neuron model the Leaky Integrate-and-Fire with Latency (LIFL)-based spiking neuron. This neuron model allows encoding more information than the common Integrate-and-Fire models typically considered for neuromorphic implementations. Moreover, the same author has been presented a novel technique in order to drive the memristive synapse.
%The proposed system employs the first circuit implementation of LIFL. Further due to the high density of connection between the neurons made by the synapses in a neural network, an improvement of the memristor based synapse circuit, just presented before by the same authors, in terms of power is presented.
%Firstly, the design of each functional block is described, and the related operating principle is compared to the biological case. Finally, in order to show how the overall system works, a motif composed of three neurons and two synapses is designed and simulated using PSpice.Today a remarkable interest to exploit the outstanding brain working due to a capacity of the brain to process data in a highly distributed fashion, requiring relative low-power consumption.
\end{abstract}


%\begin{keyword}
%\texttt{elsarticle.cls}\sep \LaTeX\sep Elsevier \sep template
%\MSC[2010] 00-01\sep  99-00
%\end{keyword}

\end{frontmatter}

%\linenumbers

\section{Introduction}

In recent years, many efforts have been done in order to reproduce the brain behaviour; this is due to a remarkable capacity of the brain itself to process data in a highly distributed fashion, requiring relative low-power consumption ($\sim$12  W \cite{Sarpeshkar1998}). Moreover, real-world stimuli are transmitted and processed with high precision within millisecond timescale. Thus, it should not be surprising that a huge number of studies have been made over the years to understand and reproduce the brain operations, involving several scientific application areas: for example, real-world data classification \cite{BohteKok2002a, BohteKok2002b, Belatreche2003}, image recognition \cite{Thorpe2001, Guyonneau2004, Perrinet2004}, speech recognition \cite{Hopfield2000, Verstraeten2005, Gutig2009}, decision making \cite{Glackin2008}, and rehabilitation \cite{Rom2007} (for an extensive review see Ponulak \emph{et al.} \cite{Ponulak2011}). 

Recently, many studies have focused their attention to understand and mimic brain-\emph{like} behaviors via hardware, leading to a class of system called \emph{neuromorphic systems} \cite{Maed1989}. The growing interest in neuromorphic systems is due to wide  range of attractive applications, such as real-time and low power spike-based computing system implementations, compact microelectronic brain-machine interfaces, among others \cite{Indiveri2011}. 

In our previous works \cite{Salerno2011}, \cite{Cardarilli2013}, \cite{Cristini2015}, we have introduced a neuron model called Leaky Integrate-and-Fire with Latency (LIFL). In this model, the spike latency phenomenon has a primary role in the spike generation process. The spike latency phenomenon has been studied before and used in various applications discussed and described in \cite{FitzHugh1955}, \cite{Izhikevich2007}, \cite{Zheng2009}, \cite{Izhikevich2010}, \cite{Chen2011}, \cite{Yamani2012}, \cite{Gollisch1108}, \cite{Fontaine2009}. 
%Hence, the hardware implementation represents a crucial aspect \cite{Qiao2015}, \cite{Stoop2013}, \cite{Cauwenberghs2009}, \cite{Saighi2011}, \cite{Sejnowski2011}, \cite{Parent2009}.

In biological systems, the synapses are tens of thousands times greater in number than neurons therefore, optimization of electronic synapse design plays a key role. Scaling of synapse area and power consumption are critical issues for designing a large scale neuromorphic circuit. However, the advancement in nanotechnology allows us to exploit today's nanodevices, such as \emph{memristor}, to save both silicon area and power consumption \cite{Strukov2008}. Despite memristor was already theorized in 1971 by Leon O. Chua \cite{Chua1971}, its ``first'' physical  implementation took place in HP Labs only in 2008 \cite{Strukov2008}. A memristor represents a two-port passive element with a charge-dependent resistance, also known as memristance. Under proper conditions, this property makes the memristor able to mimic not only the STDP (i.e., Spike-Timing-Dependent Plasticity) behavior but also to enrich the synaptic dynamics by shaping the STDP learning window \cite{Serrano2013}. The STDP mechanism is well accepted by the scientific community as a process underlying learning and memory in the brain (for a detailed overview, see \cite{Bi1998},\cite{Markram2012}).
Thus, many studies have been made to exploit memristors with the purpose of realizing plastic synapses within neuromorphic systems \cite{ Serrano2013, ZamarrenoRamos2011, Indiveri2013, Perchin2014, Bill2014, Guo2014, He2014, Lecerf2014, Wu2015}. In these works shape-tailored waveform pulses are often employed, limiting the choice of the neuron model to be used (usually, I$\&$F \cite{Abbott1999} models type).

To mitigate this aspect, the authors have presented in \cite{prime2016} a novel circuit implementation of a memristor-based synapse able to work with any arbitrary pulse shape. Furthermore, in \cite{el} the authors have presented a PCB circuit emulating the biological spike firing scheme that activates the memristor synapse. 
\\In this work, we present a hardware implementation of the complete LIFL neuron model. Further, an improved  synapse driving circuit is presented.
The paper is organized as follow: in section \ref{TM} we give the theoretical model of LIFL and synapse. A brief review of memristor device is given in section \ref{mem}. In section \ref{CI} we describe the circuit implementation for the neuron and the synapse, followed by simulation results and discussion of a simple motif in section \ref{discuss}, and finally conclusion of the work.
%The paper is organized as follow: in section \ref{neuro} and section \ref{e} we give the theoretical model of LIFL and synapse respectively. A brief review of memristor device is given in section \ref{mem}. In the following sections \ref{N_s}, \ref{sy_s} we describe the circuit implementation for the neuron and the synapse, followed by result and discussion of a simple motif in section \ref{discuss}, and finally conclusion of the work.}}

\section{Theoretical models}
\label{TM}
\subsection{Neuron} 
\label{neuro}

In this section we considered spiking neural model  with the aim of modeling a silicon neuron. The used model is the LIFL (Leaky Integrate-and-Fire with Latency) \cite{Cardarilli2013}, \cite{Cristini2015}. It is characterized by two different behaviors, sub- and suprathreshold, depending if the inner state ($S$, i.e. the membrane potential) is under or over a spiking threshold ($S_{th}$), respectively. In subthreshold ($S$ $< S_{th}$, a.k.a. passive mode) the model exhibits a leaky integrator behavior; whereas, in suprathreshold ($S$ $\ge S_{th}$,  a.k.a. active mode), a neuron does not fire instantaneously, but after a continuous-time delay called \emph{time-to-fire, $t_f$} (i.e., the spike latency of the biological counterpart). The spike latency mechanism allows the strength of the input to be encoded by the spike timing, as observed in most cortical neurons \cite{Izhikevich2004}. The models with a different behavior (i.e., I$\&$F models) show a lack of information and are not bio-realistic \cite{Izhikevich2007}.

The neuron model is described by the following  equations:

\begin{eqnarray}
S=S_p+P_rP_w-L_d \Delta t \; , \mbox{ if } S < S_{th}   \label{pm}
\\
S=S_p+P_rP_w+\frac{(S_{p}-1)^{2} \Delta t}{1-(S_{p}-1)\Delta t} \; , \mbox{ if } S \ge S_{th}\label{eq:am}
\end{eqnarray}
\label{eqtot}
\begin{equation}
t_f = \frac{1}{(S-1)} \; \label{eq:fe}
\end{equation}

\begin{figure}[!ht]
\centering
\includegraphics[width=0.9\textwidth]{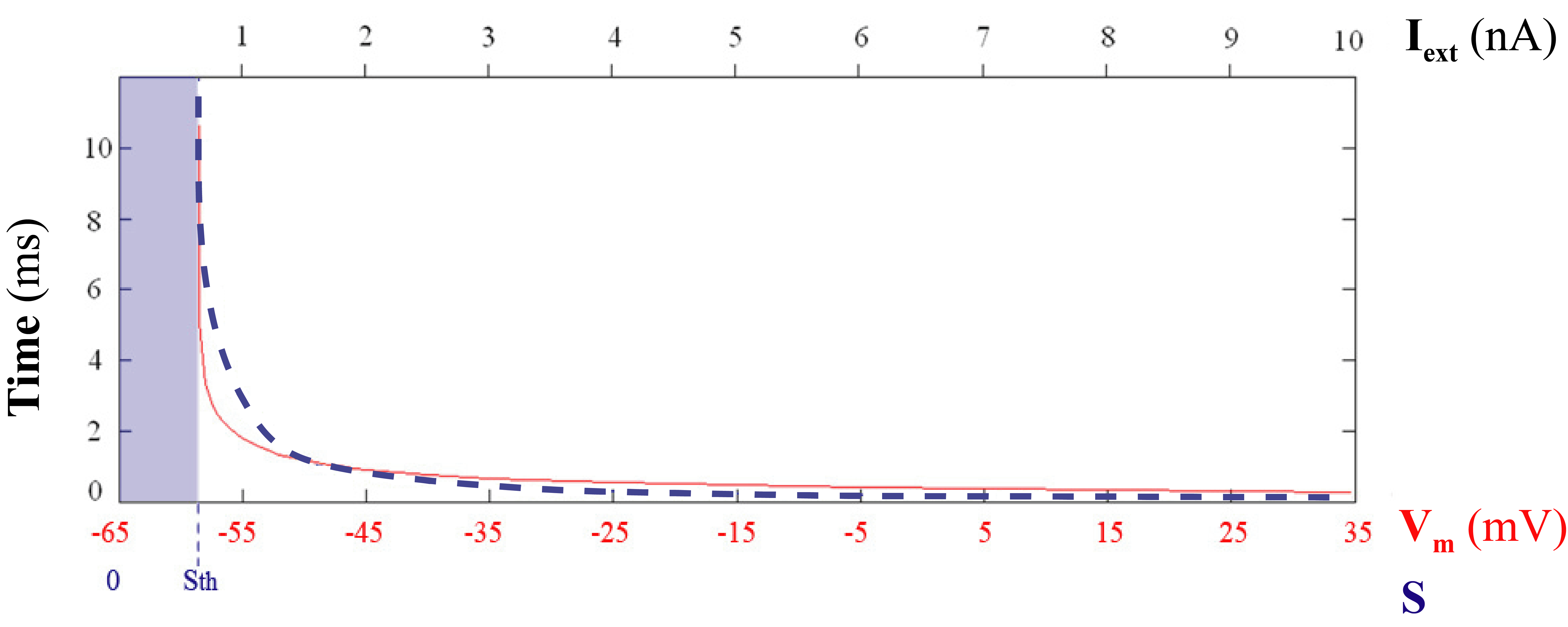}
 \caption{Comparison between Latency (red) and \emph{firing equation} (dotted blue). Modified from \cite{Cristini2015}.
 }
\label{F:comparison_latency_firing_equation}
\end{figure}

In these equations, the state $S$ is in the range [0, $\infty$) (where the lower bound represents the resting state), $S_p$ is the previous state and $P_r$ is the ``presynaptic weight'' (or spike), representing the output generated by a firing neuron (ideally a Dirac delta). During the propagation towards the target neuron, this quantity is multiplied by $P_w$, ``postsynaptic weight'', bounded in the range [0, 1]. The $S_{th}$ is the spiking threshold, expressed by $1+d$, where $d$ is chosen in order to have a maximum finite spike latency as in biological systems \cite{FitzHugh1955}. Indeed, for $S = S_{th} = 1+d$, the time-to-fire calculated through Eq. \ref{eq:fe} is $t_{f,max}=1/d$. Moreover, $L_d$ is a positive quantity representing the linear subthreshold decay (note that when $L_d = 0$ the neuron behaves like a \emph{perfect integrator}), $\Delta t$ is the temporal distance between two consecutive incoming spikes.
The Eq. \ref{eq:fe} represents an approximation of the curve obtained through the simulation of a membrane patch stimulated by brief current pulses (0.001 ms of duration). This approximation was done solving the non linear and differential Hodgkin-Huxley equations \cite{Hodgkin1939}, \cite{Hodgkin1952} and using NEURON simulator \cite{hines1997neuron}. We called this relationship \emph{firing equation}. In Fig. \ref{F:comparison_latency_firing_equation}, a qualitative comparison between the simulated behavior of the latency and the firing equation is shown.

In order to be effective, an incoming pre-synaptic spike must occur before the spike generation. Then, the denominator of the fractional term in Eq. \ref{eq:am} is always greater than zero. The fractional term allows the evaluation of the inner state of neuron target when it receives further inputs during the $t_f$ time window. Finally, after the firing, the neuron is reset to its resting potential (i.e., $ S = 0$) for a time equal to $t_{arp}$ (i.e., absolute refractory period), in which the neuron remains insensitive to further incoming spikes.

The continuous-time behavior of the model is suitable for event-driven simulation methods, by which it is possible to save overall information \cite{Brette2007}. For a detailed description of the model see \cite{Cristini2015}.

\subsection{Synapse and STDP} 
\label{e}
\begin{figure}[!ht]
\centering
\includegraphics[width=0.5\textwidth]{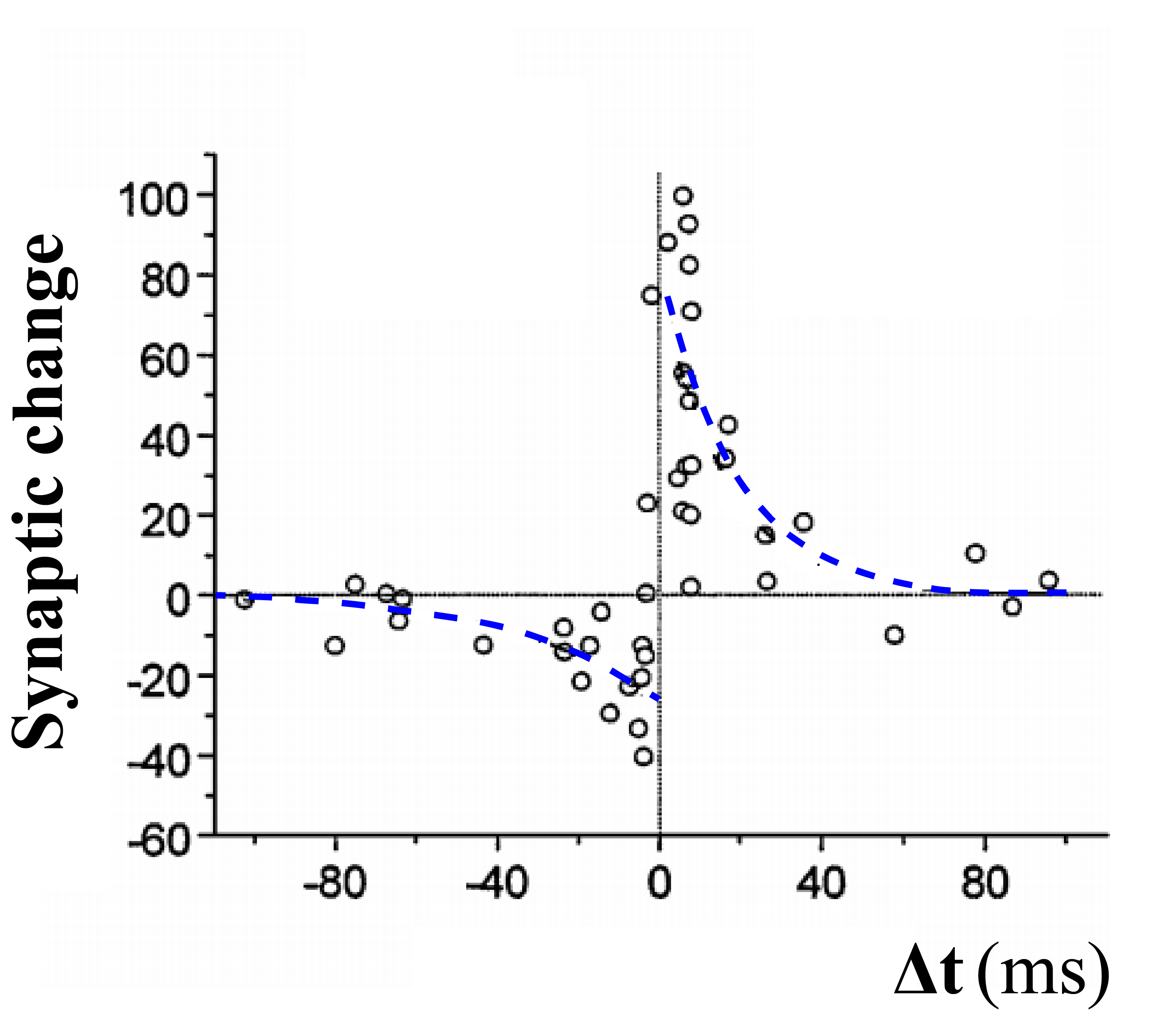}
 \caption{STDP behavior. Long Term Depression LTD $(\Delta$t $<$ 0).
 Long Term Potentiation LTD ($\Delta$t $>$ 0). 
 Modified by \cite{Bi1998} }
\label{F:STDP1}
\end{figure}

\begin{figure}[!ht]
\centering
\includegraphics[width=0.6\textwidth]{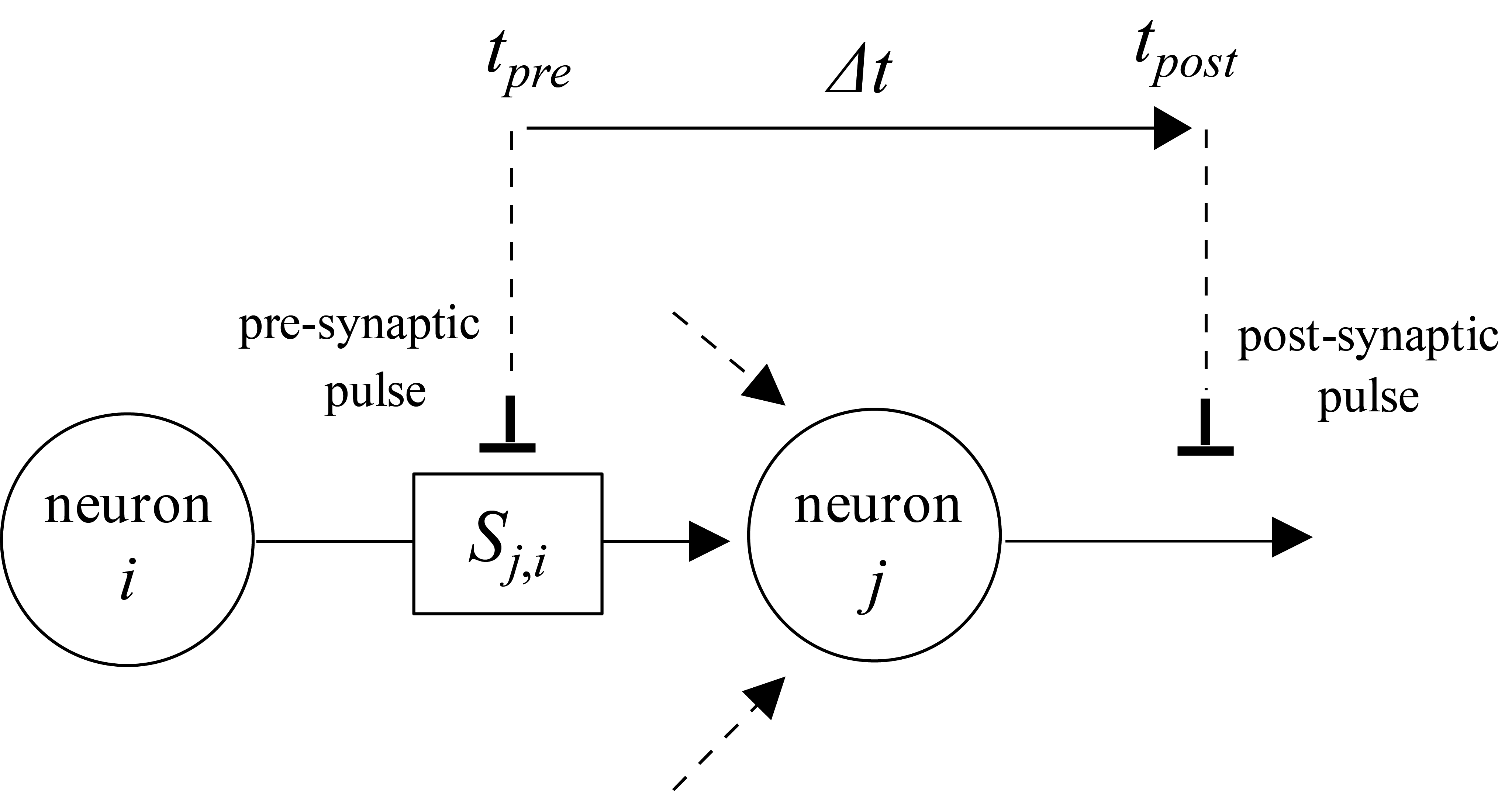}
 \caption{ $\Delta$t representation.}
\label{F:delta}
\end{figure}

Neurons in the network are connected by links (i.e., synapses) characterized by synaptic weights which permits to modify the amplitude of the passing pulses. Synapses can vary according to the activity of the network.
Synaptic plasticity is the ability of synapses to strengthen or weaken of the weight over time, in response to increases or decreases in their activity. A well-known type of synaptic plasticity is based on the precise timings of pre- and post-synaptic spikes, influencing the magnitude and direction of change of the synaptic strength \cite{Sinha}.
This rule is referred to as Spike-Timing-Dependent Plasticity (STDP, see \cite{Bi1998}), found by Bi \& Poo through an experimental protocol. In order to show the general synaptic change trend, a properly elaborated STDP behavior is represented in Fig. \ref{F:STDP1}. The change of synaptic weight is plotted as a function of the relative timing between pre-synaptic spike arrival and post-synaptic firing. The right dashed blue curve represents the change of synaptic weight when pre-synaptic spike arrives before the post-synaptic spike is generated; the left curve dashed blue represents the change of synaptic weight when post-synaptic spike is generated before the pre-synaptic spike is arrived. In Fig. \ref{F:delta}  $\Delta$t is represented, with reference to the synapse $S_{j,i}$. Note that  $\Delta$t = $t_{post} - t_{pre}$.

The STDP behavior can be approximated by Eq. \ref{eq:case}, which gives the modification in synaptic strength \emph{i.e. synaptic change} for a pre-synaptic arrival time, $t_{pre}$, and a post-synaptic firing time, $t_{post}$:
% $\frac{\Delta w_{j,i}}{w_{j,i}}$

\begin{equation}
\label{eq:case}
synaptic\;change =\begin {cases}
%\frac{\Delta w_{j,i}}{w_{j,i}}=\begin{cases}
 A_{-}exp \frac{\Delta t}{\tau^{-}} & \text{for $\Delta$t $<$ 0,}\\
0 & \text{for $\Delta$t = 0,}\\
A_{+}exp \frac{-\Delta t}{\tau^{+}} & \text{for $\Delta$t $>$ 0,}\\
\end{cases}
\end{equation}

In the  above equation (Eq.\ref{eq:case}), $\tau^{+}$ and $\tau^{-}$ are the time constants for potentiation and depression, respectively, and $A_{+}$ and $A_{-}$ are the maximum amplitudes of potentiation and depression, respectively. All these parameters are positive, except $A_{-}$.

Although STDP varies tremendously across synapse types and brain regions \cite{Abbott2000}, a standard choice for these parameters, obtained by fitting the real case \cite{Viriyopase}, is shown in Table \ref{table:value}.

\begin{table}[]
\centering
\caption{standard values set for STDP.}
\label{table:value}
\begin{tabular}{|l|}
 \hline
$\tau$\textsuperscript{+}= 16.8 ms\\  \hline
$\tau$\textsuperscript{-}= 33.7ms  \\  \hline
A$_+$ = 0.78  \\  \hline
 A$_-$=  - 0.27   \\ \hline
 \end{tabular}
\end{table}

Note that in some variants of STDP models $A_{+}$ and $A_{-}$ are represented by proper functions in order to keep the synaptic weights in a bio-plausible range of variation. 

Although experimental studies usually report synaptic change as a fractional change $(i.e.,   \frac{\Delta w_{j,i}}{w_{j,i}})$ resulting after some number of pairings. In many models it's assumed that changes induced by spike pairings at a particular $\Delta$t are absolute changes with units of conductance (e.g.\cite{Farries2007}). In this work the latter choice is adopted.

\section{Memristor}
\label{mem}
Memristor is a two terminal device originally postulated by L.O. Chua in 1971 \cite{Chua1971}.
%As described by Chua, memristor would necessarily show a firm static relationship between the charge ($q$) flowing through the device and flux ($\Phi $). If the relation between the $q$ and $\Phi $ is linear the memristor behaves as a linear resistor.  
%Essentially, memristors are variable resistors with the ability to store/memorize their previous resistive value/state, 
The resistance of memristor (memristance) can be controlled by changing the input that can be a voltage or a current, in this manner we have voltage/flux controlled or current/charge controlled memristor. The voltage controlled memristor can be described by the following equations:    \par

\begin{gather}
I_{MR} =  M_G (w, V_{MR}, t) V_{MR}
\\
w= f(w, V_{MR}, t)
\end{gather}

Where, \textit{$ I_{MR} $, $V_{MR}$} are the current and the voltage input to the memristor, respectively. \textit {$M_G$}, \textit{$ M_R$} represent the memconductance  and the memristance, respectively \cite{Chua1976}. The current controlled memristor can be described by the following equations: \par

\begin{gather}
V_{MR} =  M_R (w,  I_{MR}, t)  I_{MR}    
\\
w = f(w, I_{MR}, t)
\end{gather}
Note that \textit {w} is the internal state parameter of the memristor.
The most important characteristic of the memristor is the pinched hysteresis loop (i.e., hysteresis loop passed through origin) in its current-voltage ($I-V$) curve when activated by an alternating signal. 
%The switching could be unipolar or bipolar depending upon material \cite{DiVentra2009}.Memristor pinched hysteresis loop ( i.e. hysteresis loop passed through origin), variable conductance with respect to change in input signal magnitude or polarity and the ability to store the last conductance state after the device is switched off  
 This characteristic is able to modify the conductance respect to the input signal change both in  magnitude and polarity, and store the last conductance state even after the device is switched off. These properties are also closely followed by the biological synapse \cite{Li2014}, \cite{Ohno2011} and in this sense, a memristor behaves like a synapse. A synapse is a connection between two neurons in the brain with a plastic/programmable synaptic weight that can be modified to alter the efficiency/strength of signal transmission between neurons under the influence of transmission itself. For this reason, several memristor based synapse implementations have been proposed in literature  \cite{Indiveri2011}, \cite{Bi1998}, \cite{Serrano2013}, \cite{ZamarrenoRamos2011}, \cite{Wu2015},  \cite{Serrano2012}, \cite{Snider2008}.\\
In this work we use a particular memristor model presented in \cite{Yakopcic2011} for simulating the proposed system. 
The Spice model of the memristor used in this experiment is given in  \cite{Yakopcic2011}. The model is very general and has good flexibility to accommodate a different type of memristor dynamics. Here, we briefly discuss the core equations used by the authors in \cite{Yakopcic2011} to developed the model. Values of all the variables used in our experiment are listed in Table \ref{table:mem}

The $I-V$ equation for the used memristor is given in Eq. \ref{equation1}  
\begin {equation}
\label{equation1}
\it \begin{cases} a_1\times x(t)\times \sinh(bV(t)) & \it V(t) \geq 0\\ a2\times x(t) \times \sinh(bV(t)) &  V(t) < 0\end{cases}\\
\end{equation}

where, $x(t)$ is memristor state variable, which defines the change in resistive state of the memristor device based on the dynamics of the particular device. In the used memristor model the state variable can have any value between 0 and 1. Where,

 \textit{ $\begin{cases} Low resistive state  & x(t) = 1\\ High resistive state & x(t) = 0\end{cases}$\\}

The parameters $a_{1}$, $a_{2}$ and $b$ are fitting parameters and their respective values used in our work is given in the Table \ref{table:mem}.
The rate of change of the state $x(t)$ is given in Eq.\ref{equation2}. The change in the state variable depends on two different functions, first $g(V(t))$ and second $f(x(t))$. Function $g(V(t))$ implements the threshold voltage limit for the model used. The second function $f(x(t))$ implements the non linear ion motion also commonly known as non linear dopant drift in the state variable motion. $\eta$ represents the direction of motion of the state variable with respect to the polarity of input voltage. $\eta$ could be $1$ or $-1$ , when $\eta$=$1$ positive voltage above the threshold will increase the value of state variable whereas when $\eta$=$-1$ positive voltage will decrease the value of state variable.

\begin{equation}
\label{equation2}
\frac {d(x)}{d(t)} \ = \eta* g(V(t))*f(x(t))
\end{equation}

Voltage threshold function for the used model is given by the following Eq.\ref{equation3}.

\begin{equation}
g(V(t))= \begin{cases} A_p (\exp V(t)-\exp V_p) & V(t) > V_p\\ - A_n (\exp(-V(t))-\exp V_n) &  V(t) < - V_n \\ 0 & -V_n \leq V(t) \leq V_p \end{cases}\\
\end{equation}
\label{equation3}
Where, $V_{p}$ and $V_{n}$ are positive and negative voltage threshold values and $A_{p}$ and $A_{n}$ are parameters that defines how fast the device changes its states after the threshold is reached. The values of $V_{p}$, $V_{n}$, $A_{p}$ and $A_{n}$ are listed in the Table \ref{table:mem}.
          
The non linear dopant drift  function $f(x(t))$ for the used model is given by the Eq.\ref{equation4}. Variables $x_{p}$ and $x_{n}$ are the points up until which the state variable motion is linear/constant. After this point the state variable motion is govern by a decaying exponential function with decaying rate of $u_{p}$ or $u_{n}$ respectively for point after $x_{p}$ or $x_{n}$. The values of these parameters used in our experiment are listed in the Table \ref{table:mem}. The variables $w_{p}$ and $w_{n}$ are the window functions used to ensure that the function $f(x(t)$ remains within the valid interval [1-0]. The windowing functions are given in Eq.\ref{equation5} and Eq. \ref{equation6}.

\begin {equation}
\label{equation4}
f(x(t))=  \begin{cases} \exp (-u_p) (x-x_p) * (w_p (x,x_p)) & x\geq x_p\\ 1 & x<x_p\\ \exp(u_n)(x+x_n-1) * (w_n(x,x_n)) & x \leq -x_n\\ 1 & x>1-x_n \end{cases}\\
\end {equation}
\begin {equation}
\label{equation5}
w_p (x,x_p)= \frac {(x_p-x)}{(1-x_p)}\ + 1
\end {equation}

\begin {equation}
\label{equation6}
w_n(x,x_n) = \frac {x}{(1-x_n)}
\end {equation}

\begin{table}[h]
\centering
\caption{Parameters use in this model.}
\label{table:mem}
\begin{tabular}{|l|l|}
 \hline
  Parameter &  Value\\
 \hline
$A_{p}$ = $A_{n}$& 4000\\
\cline{1-2}
$x_{p}$& 0.3\\
\cline{1-2}
$x_{n}$& 0.5 \\
\cline{1-2}
$\alpha_{p}$& 1 \\
\cline{1-2}
$\alpha_{n}$& 5 \\
\cline{1-2}
$a_{1}$ = $a_{2}$ & 0.17 \\
\cline{1-2}
$b$& 0.05 \\
\cline{1-2}
%$x_{0}$& 0.11 \\
%\cline{1-2}
$V_{p}$& 0.16 [V] \\
\cline{1-2}
$V_{n}$& -0.15 [V] \\
 \hline
 \end{tabular}
\end{table}

%Thus, memristor has the potential to generate very compact and precise model of synapse, just as representing a ``potentiated'' or ``depressed'' state by its conductance change with respect to the input. In this regard, memristors can be integrated into a highly dense crossbar network array to connect with a high number of CMOS neurons or silicon neurons and in order to emulate spike-based learning. It has also been demonstrated by some authors, that it is possible to obtain biologically plausible STDP weight change and shape the form of STDP learning function by shaping both pre- and postsynaptic spikes \cite{likharev2011crossnets}, 

\section{Circuit implementation}
\label{CI}
In this section, we start to describe the overall system, composed of the neuron and synapse models. All the schematics and the simulations are performed in PSpice environment. 
\subsection{Neuron} 
\label{N_s} 
\begin{figure}[!htbp]
\centering
\includegraphics[width=1\textwidth]{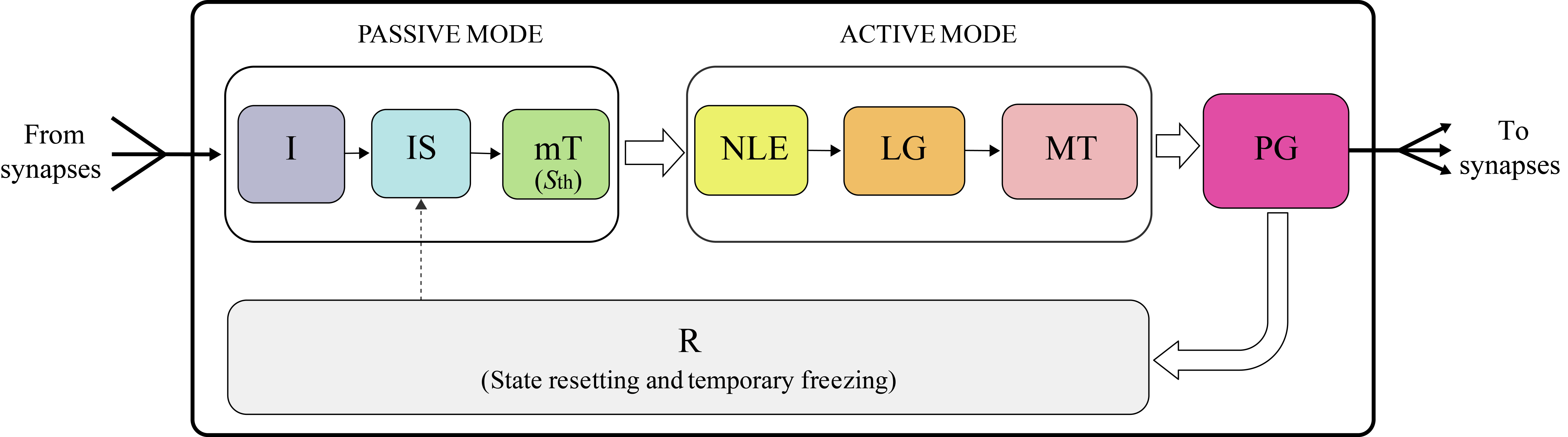}
\caption{Logical Block of the overall circuit. I: integrator block; IS: internal state block; mT: mimimum threshold block; NLE: non-linear element block; LG: latency generation block; MT: maximum threshold block; PG: pulse generator block; R: refractory block.}
%\vspace{-0.1cm}
\label{Neuron Logic Diagram}
\end{figure}

Here we illustrate the circuit implementation of the neuron model.
As shown in Fig. \ref{Neuron Logic Diagram}, it is composed of a certain number of logical sub-systems:

\begin{enumerate}
\item Integrator circuit (I): composed of the input stage of an OTA.
\item Internal State (IS): it consists of RC group.
\item Minimum Threshold (mT): an input stage of a differential amplifier.
\item Non Linear Element (NLE): common-source amplifiers, peak detectors (diode and capacitor), shunt, and voltage translators. 
\item Latency Generation (LG): a ramp generator, a monostable and an adder.
\item Maximum Threshold (MT): input stage of a differential amplifier. 
\item Pulse Generator (PG): a monostable and an output buffer.
\item Refractory (R): a peak detector with loss (diode and capacitor and resistor) and a buffer.
\end{enumerate}

%PER LUCA
%In the following we give a schematic 
%
%Integrator circuit:(Dark purple) composed by the input stage of an OTA.
%
%Internal state: (Light blue) consists of RC.
%
%Threshold minimum:(Dark green) differential amplifier only the input stage.
%
%Non linear element: (Ocher yellow) common-source amplifiers, peak detectors (diode and capacitor), shunt, and voltage translators.
%
%Latency generation: (Dark orange) ramp generator, monostable and adder.
%
%Threshold maximum : (Pink) differential amplifier only the input stage.
%
%Pulse generator: (Dark pink) monostable and output buffer.
% 
%Refractory: (Gray) buffer and peak detector with loss (diode and capacitor and resistor). 
%
%Decrizione funzionamento (non so se è bene darla anche con qualche grafico di pspice)

The functionality of the circuit is described the case of a pre-synaptic spike (or pulse) able (or not) to trigger a post-synaptic spike.
%\begin{enumerate}[label=(\roman*)]
%\item Case of a pre-synaptic spike (or pulse) able (or not) to trigger a post-synaptic spike.
%\item Case of two consecutive pre-synaptic pulses able to change the post-synaptic pulse timing.
%\end{enumerate}

In this case, when the input pulse arrives, the value of voltage of the \emph{internal state} (light blue block \ref{Neuron Logic Diagram})  instantaneously changes. If this value is over the threshold (i.e., over the minimum threshold), the \emph{latency generation} block is triggered. When the latter reaches the maximum value, a one shot component (dark pink block in Fig. \ref{Neuron Logic Diagram}) generates a pulse (i.e., the post-synaptic spike). Moreover, at the same time the \emph{refractory} block (gray in Fig. \ref{Neuron Logic Diagram}) will be activated and other pulses will have no effect to the internal state. Whereas, if the internal state is under the threshold, it will discharge and no post-synaptic spike will be generated.
As described in section \ref{eqtot}, the neuron has two behavioral modes, passive and active described by  Eq. \ref{pm} and  Eq. \ref{eq:am} respectively. The minimum internal state threshold ($S_{th}$) divides the active mode from passive mode  and this is the first threshold in our model (green block in Fig.\ref{Neuron Logic Diagram}). When in active mode, the neuron does not fire instantaneously, but waits for an certain amount of time called time to fire ($t_f$) (see Eq. \ref{eq:fe}) before to fire. This is the second threshold in our model (pink block in Fig.\ref{Neuron Logic Diagram}). Therefore, these two thresholds were used in our model to accommodate $S_{th}$ and $t_f$.
 
\begin{figure}[!htbp]
\centering
\includegraphics[scale=0.4]{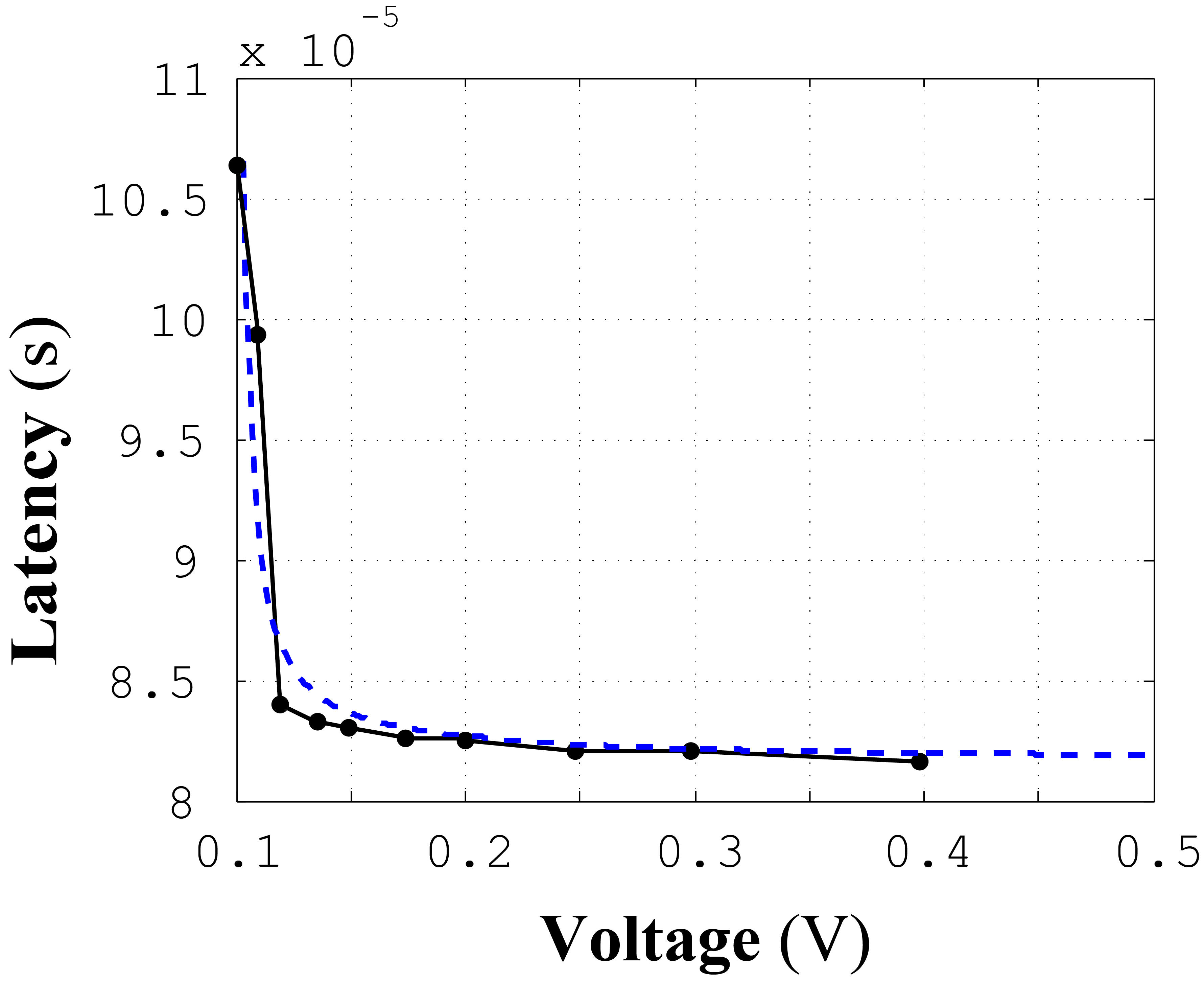}
\caption{Latency behavior simulated using PSpice (black) compared with the ideal one (dotted blue). The latter is a rectangular hyperbola which parameters are properly tuned to that of LIFL neuron model \cite{Cardarilli2013}. }
%\vspace{-0.1cm}
\label{Latency comparison}
\end{figure}

In Fig.\ref{Latency comparison} is shown the latency/input voltage characteristic of the simulated neuron, compared with the ideal one (i.e., \emph{firing equation}, see sect. 2.1). Note that we have considered the effects of latency in the input-output response for different input voltage values. From the simulation result shown in Fig.\ref{Latency comparison} , the trend approximately follows the ideal curve (i.e., a branch of rectangular hyperbola).
%"Note that we have considered the effects of latency in the input-output
%response for different input values" si può migliorare scrivendo "Note that
%we have considered the effects of latency in the input-output response for
%different input *voltage* values*, and, as we can see from Fig. 5, the
%trend **approximately **follows the ideal curve (i.e., a branch of
%rectangular hyperbola).*"

%In reference to the second case, in Fig. \ref{Latency family} is shown a family of curves obtained by changing not only the pulse amplitude of the second pre-synaptic pulse, but also the time difference between the two pulses. Note that the more is the time difference between the two pre-synaptic spikes, the less will be the effect of the second pulse on the post-synaptic response, independently of the pre-synaptic magnitude, as for the red curve in Fig. \ref{Latency family}. In the opposite case the latency of the post-synaptic pulse will be reduced (as for the blue curve in Fig. \ref{Latency family}). 
%Of course, this occurs when the second pre-synaptic pulse is excitatory. For inhibitory pulses, the latency in the generation of the post-synaptic pulse would be increased. 

\begin{figure}[!htbp]
\centering
\includegraphics[width=1.2\textwidth]{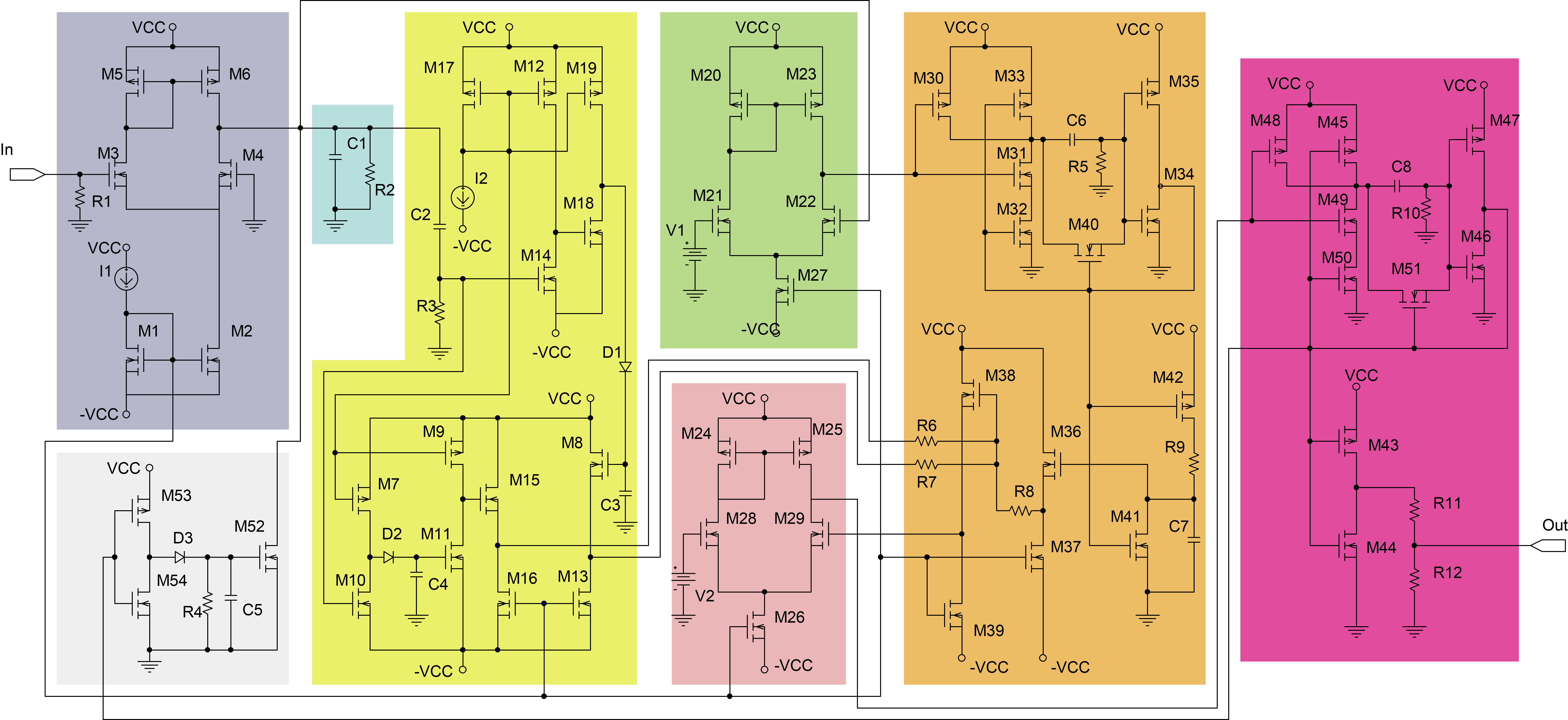}
\caption{Expanded scheme of overall neuron circuit of Fig. \ref{Neuron Logic Diagram}. The color used in this figure are coherent with those of Fig. \ref{Neuron Logic Diagram}.  }
%\vspace{-0.1cm}
\label{Neuron}
\end{figure}

In Fig. \ref{Neuron} is shown the complete PSpice model of the neuron, realized cascading the blocks shown in Fig. \ref{Neuron Logic Diagram}.

Finally, in Tab. \ref{table:res} the number of devices used in the circuit are listed. 

The resulting circuit shows low complexity associate with good bio-plausibility (for instance, like FitzHugh Nagumo model \cite{Izhikevich2004}, \cite{fitzhugh1961impulses}). The number of transistors is instead comparable with other implementations \cite{rangan2010subthreshold}, \cite{demirkol2011low}, \cite{Binczak2010}, \cite{Petrovas2012} that do not present the spike latency generation. Moreover, thanks to the implementation of subthreshold integration, refractoriness and spike latency behavior, this model mimics quite well the biological counterpart.

%%Ora farei breve descrizione per ogni blocco: 
%\subsubsection{\textit{Integrator} block} 
%Ingressi: ... Uscite: ... Funzionamento: ...
%\subsubsection{\textit{Internal state} block} 
%Ingressi: ... Uscite: ... Funzionamento: ...
%\subsubsection{\textit{Minimum threshold} block} 
%Ingressi: ... Uscite: ... Funzionamento: ...
%\subsubsection{\textit{Non-linear element} block}
%Ingressi: ... Uscite: ... Funzionamento: ... 
%\subsubsection{\textit{Latency generation} block} 
%Ingressi: ... Uscite: ... Funzionamento: ...
%\subsubsection{\textit{Maximum threshold} block} 
%Ingressi: ... Uscite: ... Funzionamento: ...
%\subsubsection{\textit{Pulse generator} block} 
%Ingressi: ... Uscite: ... Funzionamento: ...
%\subsubsection{\textit{Refractory} block} 
%Ingressi: ... Uscite: ... Funzionamento: ...

\begin{table}[]
\centering
\caption{Circuit complexity in terms of used devices for neuron.}
\label{table:res}
\begin{tabular}{|l|l|}
 \hline
  Devices&  Number\\
 \hline
 MOS transistors& 54   \\
Resistors & 12  \\
 Capacitors& 8   \\
Current sources & 2 \\
 Supply voltages& 2 \\
 \hline
 \end{tabular}
\end{table}

\subsection{Synapse and STDP}
\label{sy_s} 
\begin{figure}[!htbp]
\centering
\includegraphics[width=0.8\textwidth]{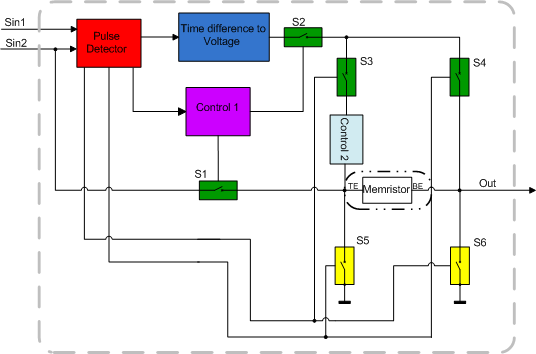}
\caption{ Logic diagram of overall system for STDP. Pulse Detector (red): it detects the pulse time order, Time difference to voltage (blue); it converts the time difference between the two pulses in an equivalent voltage, Control 1 (fuchsia); it allows the synapse weight update, S1, S2, S3, S4, switches (green), Control 2 (light blue): it mitigates the asymmetry of the memristor, S5, S6 switches (yellow).}
%\vspace{-0.1cm}
\label{STDPlg}
\end{figure}

\begin{figure}[!htbp]
\centering
\includegraphics[width=1\textwidth]{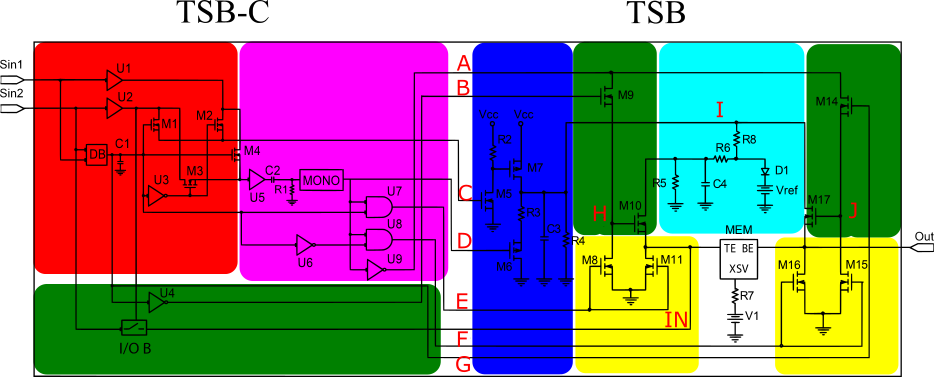}
\caption{Overall circuit scheme using the colors of Fig. \ref{STDPlg}. Sin1 and Sin2 are the input of the circuit, Out is the output of the system.}
\label{Block}
\end{figure}

The logic diagram of the STDP system is illustrated in Fig. \ref{STDPlg}. While more details about the circuit implementation are given in Fig. \ref{Block}. 
%Fig. \ref{Block}.

\begin{figure}[!htbp]
\centering
\includegraphics[width=0.8\textwidth]{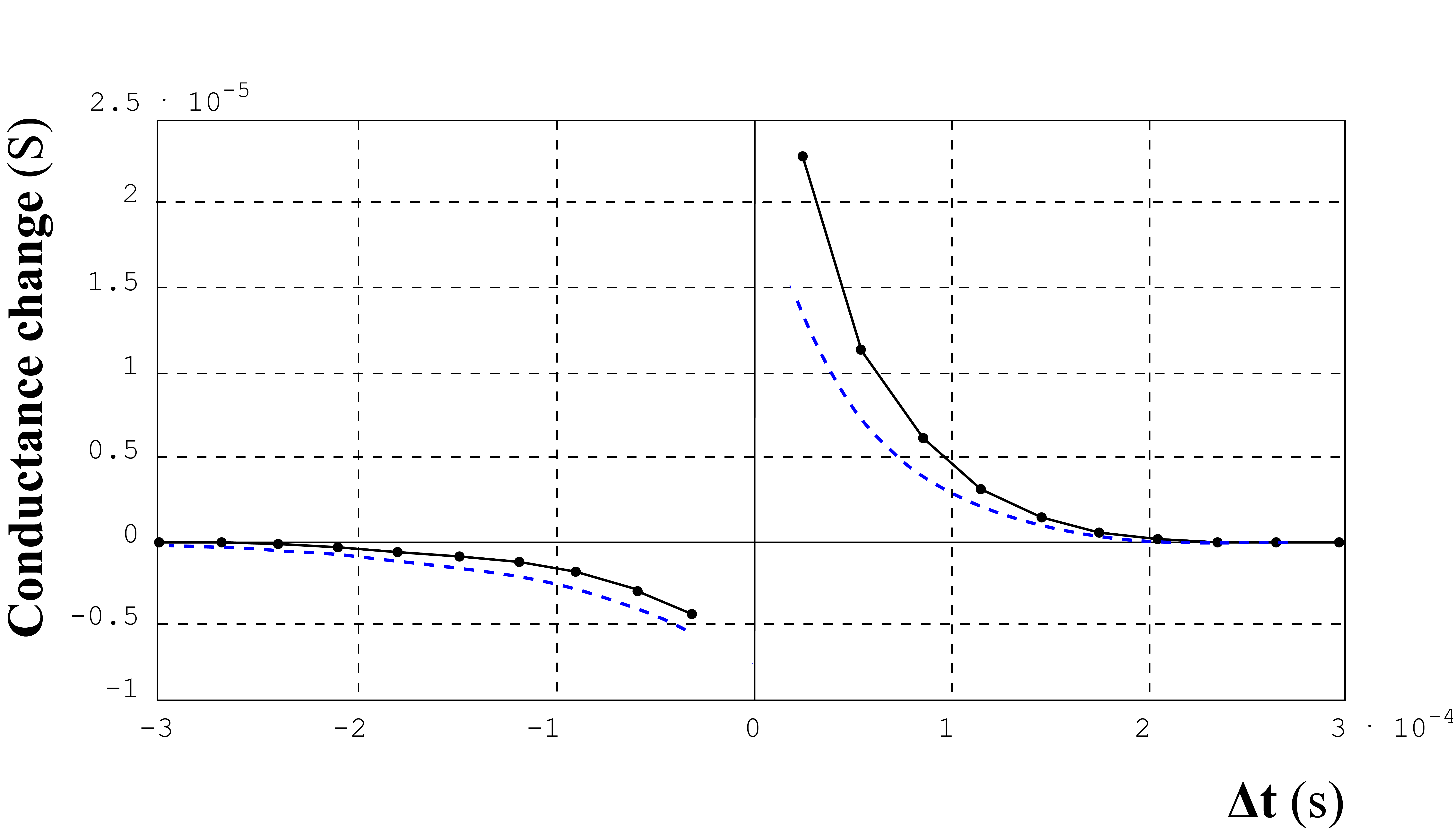}
\caption{STDP-like behavior simulated using PSpice (black) compared with the ideal one (dotted blue, see Fig. \ref{F:STDP1}) . The latter has been traced using $\frac{A_{+}}{A_{-}}$ and $\frac{tau^{+}}{tau^{-}}$ ratios obtained from standard STDP parameters, listed in table \ref{table:value}. }
%\vspace{-0.1cm} see Fig. \ref{F:STDP1}
\label{STDPb}
\end{figure}

The system of Fig. \ref{STDPlg} is composed of two main logic blocks. The first one (dashed- dot black) is the synapse (i.e., the memristor) and the second one (dashed gray) is the synapse weight update system (i.e., the system that changes the memristor conductance according to Eq. \ref{eq:case}).
In addition there are two modes of operation. 
\begin{enumerate}

\item Normal Mode: In this mode, the system works as a synapse. The signal (pre-pulse) enters through terminal Sin2 and exits out through S1 (close) and memristor. In this case, the switches S2, S3, S4, S5, S6 are open and the other blocks are off. 

\item Synapse Weight Update: In this mode, the system updates the synaptic weight (memristor conductance) according to the Spike-Timing-Dependent Plasticity rule. The pulse detector closes S3 and S6 (i.e., it  connects TE and BE terminal to the switch S2 and ground, respectively) and opens S4 and S5 allowing the increasing of memristor conductance (pre pulse arrives before post pulse). In the other case (post pulse arrives before the pre pulse), it closes S4 and S5 (i.e., it  connects BE and TE terminal to the switch S2 and ground, respectively) and opens S3 and S6, allowing the decreasing of memristor conductance.
%In this manner, we obtain a changing of conductance with the same voltage applied to the device.
The "Time difference to Voltage" block, converts the difference of time between the pre pulse and post pulse in an equivalent voltage. "Control 1" block, opens S1 and closes S2 allowing the changing of memristor conductance. Finally, "Control 2" block, mitigates the asymmetry of I-V curve of memristor.    
%(i.e., increase the conductance of memristor) (i.e., it connects TE terminal to the switch S2 )
%(i.e., decrease the conductance of memristor)(i.e., BE terminal is connected to the  switch S2 ) 
\end{enumerate}
In Fig. \ref{Block}, the circuit implementation of the above system is shown.  
Its simulation behavior is shown in Fig. \ref{STDPb}.
The circuit is composed of two basic blocks:
%The behavior of the system is the following: the circuit charges a capacitor in correspondence with the arrival of the pre-synaptic spike. Then the capacitor slowly start to discharges with an appropriate time constant. The discharge process stops when the post-synaptic spike arrives. Thus, the voltage stored is proportional to the elapsed time from pre-synaptic spike to post-synaptic spike. The behavior obtained from our circuit is shown in Fig. \ref{STDPb}.

\begin{enumerate}
\item Time Sensing Block-Control (TSB-C): This block detects the order of arrival of the two pulses (pre and post-pulses) and generates the signals for the control of the change of the mem-conductance. 
\item Time Sensing Block (TSB): This block generates a voltage proportional to the time difference of pre and post-pulses. 
\end{enumerate}

% allowing for mem-conductance changes and  and glue logic. 
%\cite{gray1990analysis} 
%\cite{baker2011cmos} 
%\cite{gray1990analysis} 

\begin{figure}[!htbp]
\centering
\includegraphics[width=0.8\textwidth]{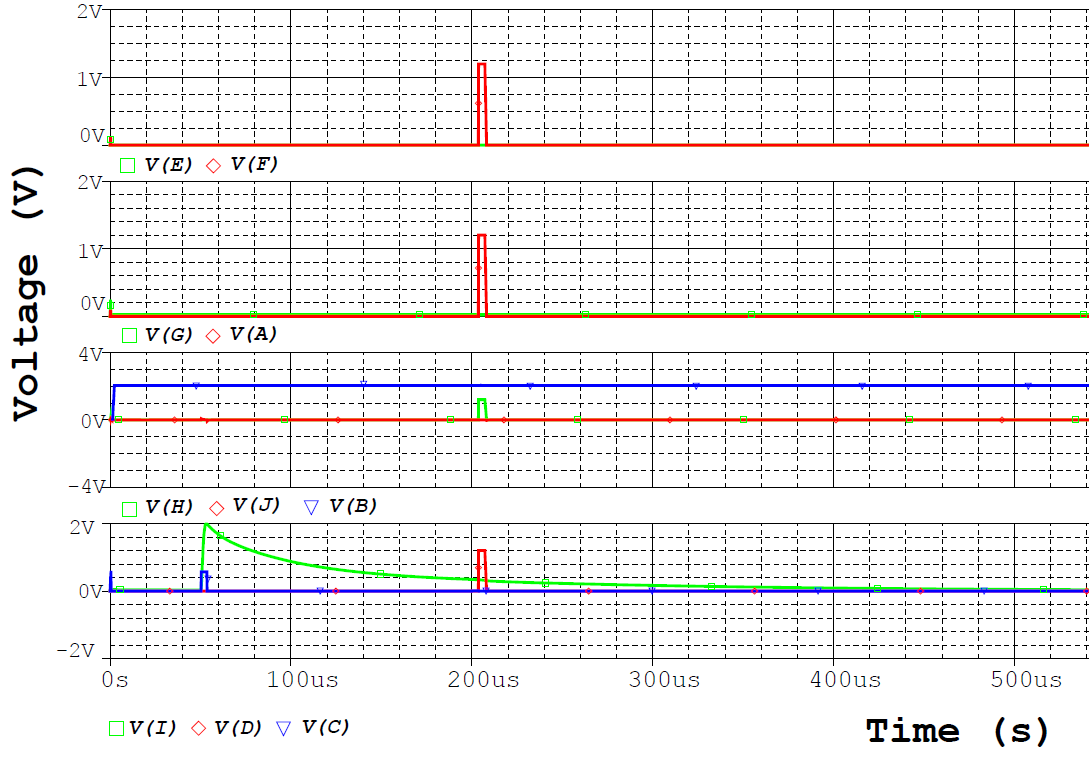}
\caption{Voltage-Time diagram for driving the TSB circuit show in Fig. \ref{Block}, for ``potentiation'' case. }
%\vspace{-0.1cm}
\label{waveform}
\end{figure}

The description of the circuit in Fig. \ref{Block} can be done taking into account the considerations developed for the scheme of Fig. \ref{STDPlg}.
NMOS transistors  M8, M9, M10, M11, M14, M15 M16 and M17 of TSB are off while the pulse enters in ``IN'' and goes through the memristor (MEM,) to ``OUT''. The increasing of the conductance of MEM is obtained by connecting the ``BE'' terminal to ground and ``TE'' to V(I) via M10. On the other hand, a decreasing of the conductance can be obtained connecting ``TE'' to ground and ``BE'' to V(I) via M17. In other words, V(B) allows the voltage stored to be transferred to the memristor.  The couple M10 and M17 connects the MEM to C3 and are controlled by the voltages V(H) and V(J). The couple M8 and M11 allows the grounding of ``TE'' terminal of MEM, whereas M16 and M15 allows the grounding of ``BE'' terminal.  V(A) is the signal for the change of mem-conductance. It is generated in TSB-C block. In this block the DB (i.e., a differential amplifier) is able to detect the arriving time of the two pulses, charging or discharging the capacitor C1, turning on M1, M4 and turning off M2, M3, when the presynaptic pulse arrives before post synaptic pulse (i.e., potentation case). In this case, when the  post pulse arrives, TRIG block (R2, C1, U5) triggers the MONO (i.e, monostable) that generates the signal V(A) (see Fig.\ref{waveform}) opening the Buffer I/O and allowing the change of memristor conductance.

%and allowing the change of memristor conductance.
%composed of level shifter allowing the signal transfer from $S_{in1}$ to IN.
 Due to the asymmetry of memristor curve, the positive conductance change (LTP) is higher than the negative one (LTD) for the same value of voltage applied to the device, Fig. \ref{STDPb}. The device set (R4, R5, R7, C2, D1, Vref) is the mitigation circuit that controls the value of voltage applied to the device for the LTP case, needed to obtain a more bio-realistic curve as in \cite{Bi1998}. Moreover, it is possible to do a tuning of the conductance value via the voltage generator Vref and supply voltage Vcc applied to transistors M5, M7. 
The MEM is the memristor-based synapse. The third terminal of memristor (``XSV'' in Fig. \ref{Block}) is only used for plotting the internal state variable \cite{Yakopcic2011}.

%and it is decribed in Fig. \ref{MEMCir}. 

%As describe in section \ref{e} the effect of the synaptic changes expressed by (\ref{eq:case}) is similar to a discharge of a capacitor.

%(B) Schematic diagram for STDP, it is formed of TSB-C and TSB. The first, (TSB-C), is composed of DB and  MONO.
%DB: differential Block able to recognize which pulse arrive first. 
%MONO: Monostable allow the change of conductance.
%The second, (TSB), is composed of T-V and V-MEM.
%T to V: Allow conversion of time in an equivalent voltage. 
%V to MEM: Applied voltage to memristor.

%In this block, STDP mechanism is basically implemented by the transistors M1, M2, M3, the capacitor C1 and the two resistors R2, R3. The other components are necessary to conduct the signal from C1 to MEM, to change the conductance value of the memristor and to allow the signal to flow from the input (``IN'') to the output (``OUT'').
%For the sake of clarity, the circuit that generates the inputs to TSB (i.e., the circuit in Fig. \ref{MEMCir}) is shown in Fig. \ref{Block}.

%In other word, the overall system can be thought as a black box that, depending on the timings of the input stream (i.e., pre- and post-pulse), is able to change the state of the memristor (i.e., its conductance). 
We remark that this circuit does not depend on the waveform of the pulses as in  \cite{ZamarrenoRamos2011}, \cite{Wu2015}, \cite{Serrano2012}, \cite{XWu2015}, but it is thought to realize a spike timing dependent circuit based on memristors, allowing a timing-sensitive behavior. The STDP behavior obtained by PSpice simulations of the whole circuit is shown in Fig. \ref{STDPb}. For further details see \cite{prime2016}. 

\begin{table}[]
\centering
\caption{Circuit complexity in terms of used devices for synapse.}
\label{table:res1}
\begin{tabular}{|l|l|}
 \hline
  Devices&  Number\\
 \hline
 MOS transistors& 68   \\
Resistors & 13  \\
 Capacitors& 6   \\
 Supply voltages& 2 \\
 \hline
 \end{tabular}
\end{table}

Tab. \ref{table:res1} gives the number of devices used in the circuit. 

\section{Simulation results and discussion}
\label{discuss}

%This section shows simulation results of a small network motif, containing three neurons (described in section \ref{N_s}) and two synapses (described in section \ref{sy_s}) is discussed.

\begin{figure}[!htbp]
\centering
\includegraphics[width=1\textwidth]{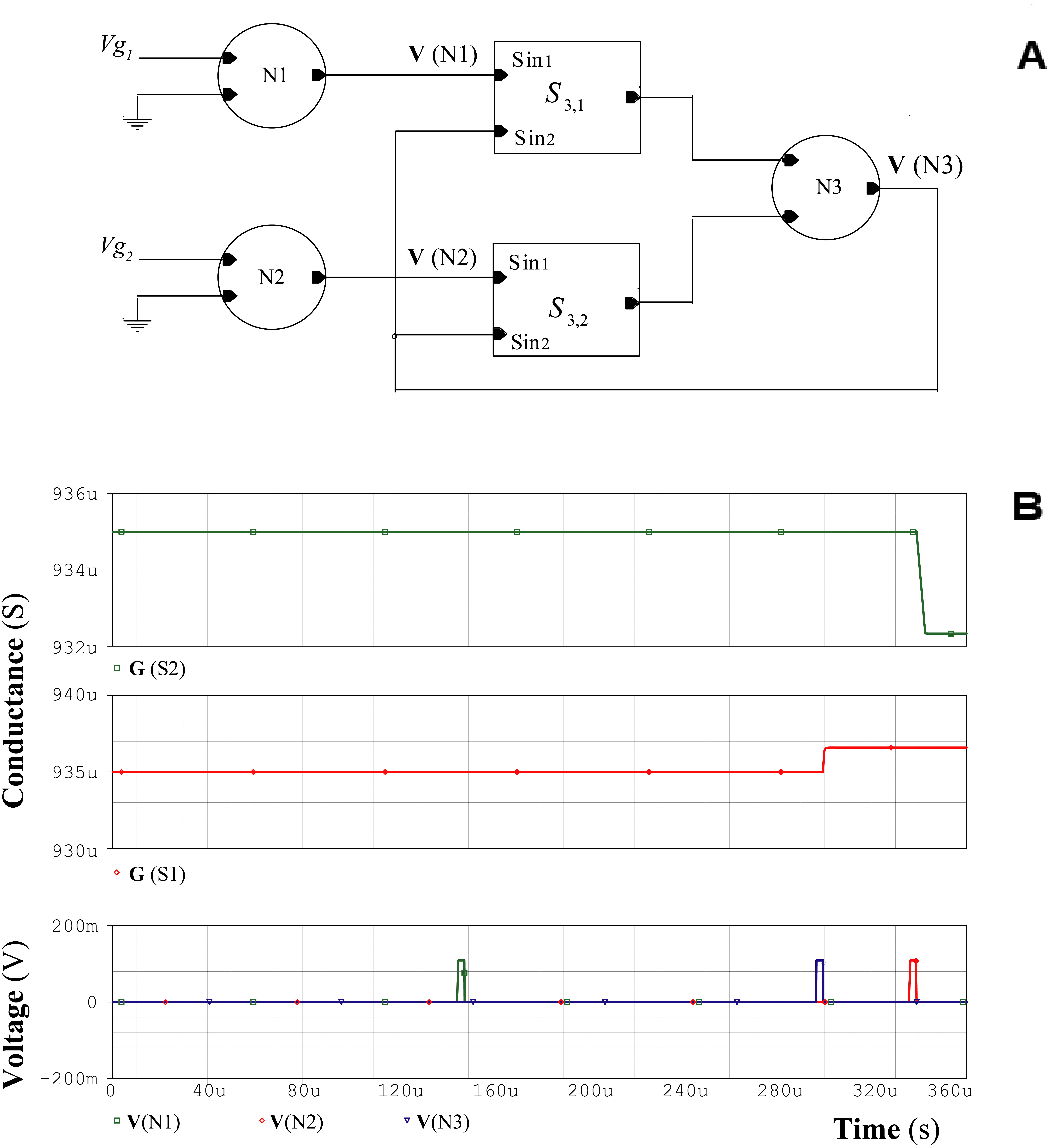}
\caption{(A) Simple simulated circuit composed of 3 neurons (N1, N2, N3) and 2 synapses (S1, S2). In order to trigger the neurons N1 and N2, two impulsive generators are necessary (i.e., $V_{g1}$ and $V_{g2}$). (B) bottom: voltage pulses; middle: LTP behavior for S1; top: LTD behavior for S2.}
%\vspace{-0.5cm}
\label{ExCir}
\end{figure}

In order to show the behavior of the overall circuits a motif system is presented. It is a neural network with two input neurons for sensory and one output neuron for the association and the decision, as shown in Fig. \ref{ExCir}A. This three neurons and two synapses system is widely used for describing and analyzing the behaviour of neural circuits \cite{Wu2015}, \cite{pershin2010},\cite{WuACMOS}. Moreover, the low complexity of the system allows to perform an electrical simulation.

%Finally, an associative learning was experimentally demonstrated by a , as shown in Fig. 8A, also known as the Pavlov’s dog [2].
% as shown in the behavior of the overall circuit is verified simulating  a typical motif system (e.g., \cite{Wu2015} \cite{pershin2010} \cite{WuACMOS})  containing three neurons (described in section \ref{N_s}) and two synapses (described in section \ref{sy_s}).
All simulations are performed using Pspice.
For both circuits we have used a CMOS 90nm 6 Metal Copper low-K level 3 technology.

%Considering the circuit illustrated in Fig. \ref{ExCir}A, we have simulated a system (e.g., \cite{Wu2015}, \cite{pershin2010})
The Fig. \ref{ExCir}(A) shows a simplified diagram of the network motif used for the simulation. The result of simulation is shown in Fig. \ref{ExCir}(B). The obtained results showed that the circuit is able to recognize the spike time order (i.e., pre-synaptic spike before post-synaptic spike, and vice versa). Further, it is able to process the time difference in an equivalent voltage, and apply this voltage value in order to change the state of memristor synapse. In this way, the LTP (i.e., Long Term Potentiation) and LTD (i.e., Long Term Depression) obtained behaviors are similar to those shown in Bi \& Poo \cite{Bi1998}.

As described in section "Synapse and STDP" (\ref{sy_s}), the proposed implementation mitigates the main limitations present in the literature:
the use of specific spike shapes \cite{ZamarrenoRamos2011}, \cite{Wu2015}, \cite{Serrano2012}, the presence of a relative low excursion for the conductance change \cite{ZamarrenoRamos2011}, \cite{XWu2015}.

In case of spiking neural networks composed of a large number of neurons, the very high number of synapses makes them the main source of power consumption.
For this reason, the synapse driving circuit presented in \cite{prime2016} has been improved taking into account the specific characteristics of our application. The most significant changes are the following.

%This result is obtained optimizing the DB, driver MONO (U3, R1, C2) and Buffer I/O blocks Fig\ref{Block}.

\begin{figure}[!h]
\centering
\includegraphics[width=0.8\textwidth]{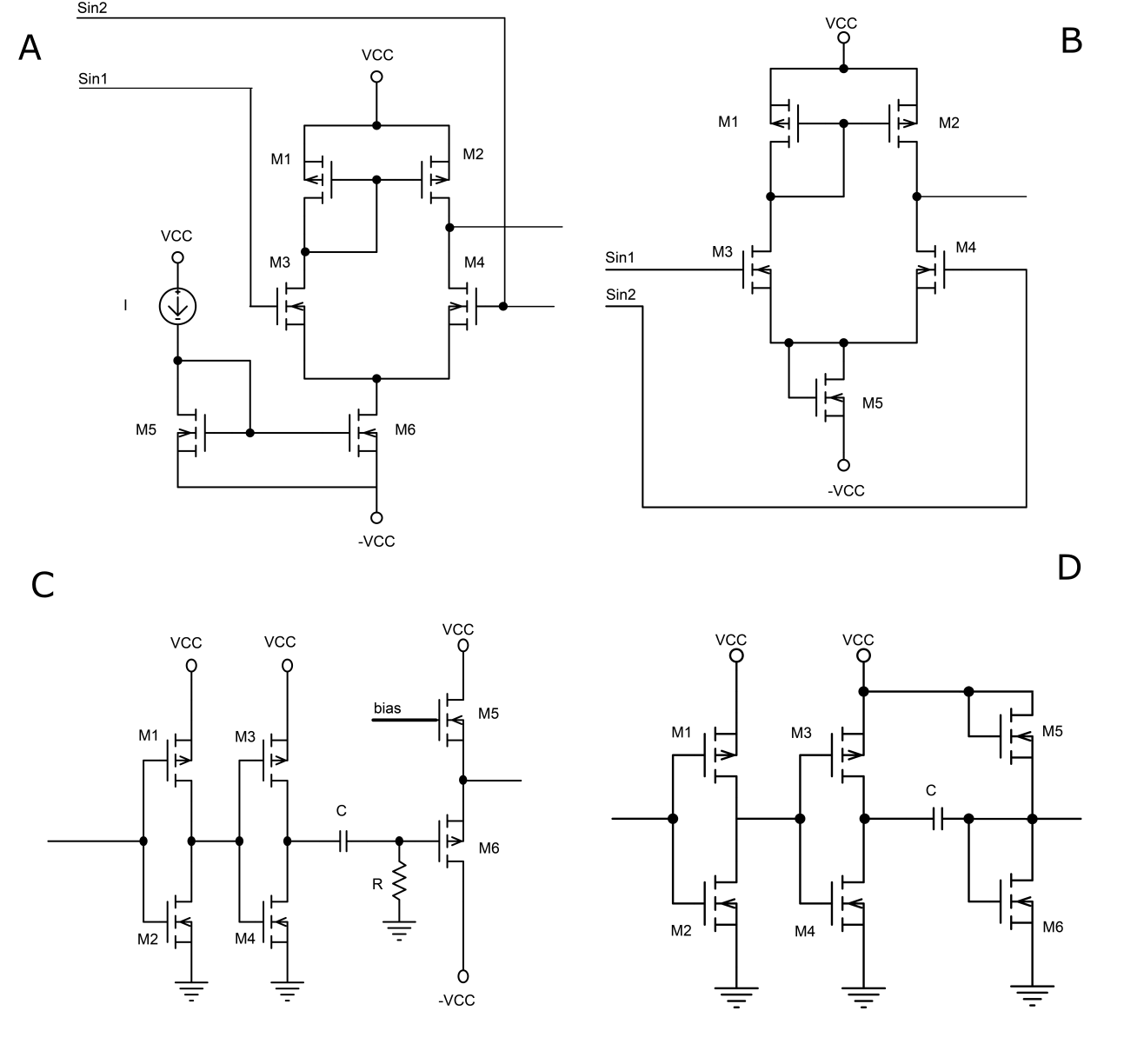}
\caption{Circuit for power reduction. Subfigure a), c) are the previous version presented in \cite{prime2016}. The other subfigure b), d) are the new ones proposed in this work.}
%\vspace{-0.1cm}
\label{switch}
\end{figure}

\begin{enumerate}
\item DB: In the original block (Fig.\ref{switch}A) the main source of power consumption is the current mirror (M5, M6) for the biasing of differential amplifier. Since the output can assume only two states (binary output),  we can reduce this bias current using a transistor in diode configuration (M5), as depicted in Fig.\ref{switch}B. 
%\item DB: In this block the main source of power consumption is the current mirror (M5, M6) for the biasing of differential amplifier Fig.\ref{dbold}. We have reduced the bias current with a transistor in diode connection (M5) as we can see in Fig. \ref{dbnew}. This reduce the biasing current and hence reduce the power consumption in this block.

%\begin{figure}[!htbp]
%\centering
%\includegraphics[width=0.8\textwidth]{1rev.png}
%\caption{DB: differential Block able to recognize which pulse arrive first.} 
%\label{BD_DETAIL}
%\end{figure}

\item TRIG: this circuit drives the MONO block in Fig.\ref{Block}. It represents the main source of power consumption is the circuit developed in \cite{prime2016}. In particular,  the biasing of level shifter (M5, M6) Fig.\ref{switch}C requires large current. Since it drives a monostable circuit, its output assumes two levels (trigger on, trigger off). As a consequence we can substitute the circuit formed of R,C and a level shifter (M5, M6) with a voltage divider (M5, M6) and capacitor (C) Fig.\ref{switch}D. 

%\item TRIG: The main source of power consumption is the current for the biasing of level shifter (M5, M6) Fig.\ref{monoold}. Instead of a trigger circuit formed of R,C and a level shifter (M5, M6), we have put a  In this block we use a voltage divider (R1, R2) and capacitor (C) Fig.\ref{mononew} for triggering the monostable multivibrator.
%\begin{figure}[!htbp]
%\centering
%\includegraphics[width=1\textwidth]{2rev.png}
%\caption{MONO:Monostable e trigger circuit.}
%%\vspace{-0.1cm}
%\label{monodriver}
%\end{figure}

%\item Buffer I/O: it is composed of voltage level shifter (M1, M2, M3, M4). It is always ON because of the %bias current and a buffer (M5, M6, M7, M8, M9) that allows the change of memconductance of %memristor Fig.\ref{switch}E. In the modified circuit depicted in Fig.\ref{switch}F, the NMOS M1 follow the %input when we have a pre pulse. Because of the low voltage of input pulse (pre pulse) 
%(0.15 V). 
%\item Buffer I/O is composed of voltage level shifter (M1, M2, M3, M4) that is always on because of the bias current and a buffer (M5, M6, M7, M8, M9) that allows the change of memconductance of memristor Fig.\ref{bufferioold}. In this case as we can see in Fig.\ref{bufferionew}, the NMOS M1 is the buffer and is turn on when we have a pre pulse, i.e, M2, M3 turn on.  
\end{enumerate}

The overall results  are  shown in table \ref{table:res2}.

\begin{table}[!h]
\centering
\caption{Comparison of a synapse driving circuit power consumption with previous work. }
\label{table:res2}
\begin{tabular}{|l|l|l|}
 \hline
Work &  Power & Area \\
 \hline
  \cite{prime2016} & 9.58 $mW$  & 500 $um^2$  \\
Present paper & 0.54 $mW$ & 480 $um^2$ \\

 \hline
 \end{tabular}
\end{table}

\section{Conclusion}
In this work, we have described the implementation of a neuromorphic system able to mimic some relevant bio-inspired neuron  features, such as integration, refractoriness, spike latency, and STDP-like behavior. The number of transistors used per single neuron in our design is comparable with other implementations, maintaining the basic bio-inspired behaviors. Moreover, we have shown a new method for driving memristor as synapse, independently of the spike pulse shape. Also, this driving system together with the memristor guarantees a good approximation of the synapse STDP-like behavior. Even though we have considered a simple network due to the heaviness of PSpice simulations, we have obtained good results in terms of bio-plausible characteristics. %Therefore, we believe that the presented work could be starting point for designing future neuromorphic circuit and systems.
In view of an on-chip implementation of a quite large neuromorphic system, we are currently improving the circuits related to both neuron and synapse driving circuit in order to further reduce silicon area and power consumption as well.

%In conclusion we have implemented a system that has the following novelties: 
%it has good bio-plausible and a number of transistors comparable with other implementations as discussed in the previous sections \ref{N_s}.
%A new system for driving memristor as synapse independent shape as in the section \ref{sy_s}. (BISOGNA EVIDENZIARE BENE IL NOSTRO MIGLIORAMENTO SECONDO ME ANCHE NELLE SEZIONI PRECEDENTI )
%\section{Bibliography styles}
%
%There are various bibliography styles available. You can select the style of your choice in the preamble of this document. These styles are Elsevier styles based on standard styles like Harvard and Vancouver. Please use Bib\TeX\ to generate your bibliography and include DOIs whenever available.
%
%Here are two sample references: \cite{Kandel2000}.

\section*{References}

\bibliography{mybibfile}

\end{document}